\documentstyle[epsfig]{elsart}
\begin{document}
\begin{frontmatter}
\title{ A Fiber Detector Radiation Hardness Test}

\author{J. B\"ahr},
\author{R. Nahnhauer}
{\renewcommand{\thefootnote}{\fnsymbol{footnote}}
\hspace{-0.3cm}\footnote{\small corresponding author, phone: +49 33762
    77346, Fax: +49 33762 77330,\\ \hspace*{0.3cm} e-mail: nahnhaue@ifh.de}},
\author{S. Nerreter\thanksref{wildau}},
\author{R. Shanidze\thanksref{tbilisi}}
\address{DESY Zeuthen, 15738 Zeuthen, Germany}
{\setcounter{footnote}{0}
\thanks[wildau]{at present student at Fachhochschule Wildau, Germany}
\thanks[tbilisi]{on leave from High Energy Physics Institute, Tbilisi State 
University, Georgia}}
\begin{abstract}
 An intense 146 $MeV/c$ pion beam was stopped inside a
  scintillating fiber detector made out of 12 planes with 16 pixels
  each, where every pixel consists out of $8 \times 8$ scintillating
  fibers of 500 $\mu$m diameter dense packed. 
  The detector was irradiated for 52 hours to more than 1 Mrad at its center.
  Before and directly after the irradiation the detector has been
  exposed to a particle beam to compare the corresponding light
  output. This study was continued during the following three
  months using cosmic rays. No damage was found taking into account the
  measurement errors of 5-10 $\%$. In contrast a 9 cm deep lucite degrader
  became irreversibly non-transparent in the irradiation region.

\end{abstract}
\end{frontmatter}

\section{Introduction}

 During the last years there was a wide discussion about the use of
different scintillator and fiber materials in high luminosity
environments for tracking and calorimetric applications for new
collider and fixed target experiments. Extensive radiation hardness
studies for these materials have been carried out to investigate
various quantities 
relevant for the problem. An introduction to the field is given in
\cite{lit1} whereas selected results of several groups are quoted in
\cite{lit2}-\cite{lit15}.

Due to the large number of relevant parameters, the published results
do not give a clear and conclusive picture of the subject. This
is true also for our own investigations which were connected with the
development of a scintillating fiber detector as backup solution for
the HERA-B inner tracker \cite{lit16}. Irradiation of scintillating and
clear
fibers using a gamma source and charged particles (p,e) at low and
high dose rates gave rather different results \cite{lit11},
\cite{lit12}-\cite{lit15} 

First of all the material under study is important concerning the   
basic polymer and the different dyes used to shift scintillation light
into the visible region of the spectrum. Decreases of light emission and
transparency are observed under irradiation. It is well known that Polystyrene
and PVT are much more radiation hard than PMMA and green shifting dyes
give better results than blue ones. It seems, however, that the kind of
production (e.g.polymerization time) can be important as well as   
temperature and atmospheric effects during transport, storage and 
machining \cite{lit3,lit10}.

An increase of the total dose rate normally will increase the damage
of the material under study. However, on an absolute scale results are  
different even for the same substances. Whereas damages are observed
in some cases already at some 10 krad \cite{lit9,lit10} other studies show
considerable effects only above 1 Mrad \cite{lit11,lit12}. In situ
observations seem to indicate  a change of the damage mechanism above this dose at
least for specific materials \cite{lit14}.

Very often a recovery of the light emission and transparency has been
reported after the irradiation. Recovery times of
some days \cite{lit4,lit6,lit14}, several weeks \cite{lit12} or even months 
\cite{lit9} have been observed. However, for the same material
total \cite{lit12} and no recovery \cite{lit11} is  reported.

The presence of oxygen during and after the irradiation seems to be of
particular importance for the damage and recovery of
materials. Parameters related to this question are dose rate,
surrounding atmosphere and coverage (glue). Published results are
again inconsistent. In \cite{lit2} it was measured that low dose rates in
air
produce larger damages than large ones. This was interpreted to be due
to a better possibility for oxygen diffusion in the material during 
longer times. The same was observed in \cite{lit15} for doses below 1
Mrad. At about that value however the same damage was found. Nearly no
effect was seen for low rates in \cite{lit13}.

If oxygen diffusion is important one would expect an influence if
air is kept off the material surfaces. No differences were
observed for irradiation in air and nitrogen in 
\cite{lit4,lit13}. 
In contrast larger effects for irradiation in argon \cite{lit10} and
nitrogen  \cite{lit11} were stated. Non-glued fibers were found to be
less damaged compared to glued ones \cite{lit11} opposite to the
measurements reported in \cite{lit12,lit13}.

All the above results underline that one is far away from a consistent
understanding of radiation damage mechanisms in scintillator and fiber
materials. In addition to the large field of parameters, measurements are 
difficult to perform. General problems are dose determination for
small probes, low light signals and mechanical damages of fragile 
objects during repeated measurements.

In order to estimate the radiation hardness of a detector of
preselected material one should therefore irradiate a finite size
prototype close to the later experimental conditions.
For the SCSF-78M fibers selected to build a high rate tracking
detector \cite{lit11,lit16} and a fast active target \cite{lit17} we made
such a test using 146 MeV/c pion beams at the Swiss Paul-Scherrer
Institute (PSI).

\section{Detector setup and readout}
Most of the following activities were done in close cooperation with  
the FAST-collaboration \cite{lit17}. In a common test run at PSI radiation     
hardness studies were performed as well as first proofs of ideas for
the precise measurement of the muon lifetime using a fast active   
scintillating fiber target.

Two detectors - $\mu FAST$ I and II - have been used for
radiation hardness tests. Both were produced at DESY
Zeuthen with the winding technology \cite{lit16}. A drum grooved with a pitch
of 510 $\mu$m has been used to produce detector planes consisting of 8
fiber  layers of $16 \times 8$ KURARAY SCSF-78M double clad fibers of 
500 $\rm \mu$m diameter dense packed. White acrylic paint has been taken
to glue all fibers of a plane together over a length of about 10 cm. One plane has
an effective width of 66 mm and a depth of 3.6 mm. Twelve of these planes
were glued on top of each other to build the final detector.

A photo of $\mu FAST$ I is displayed in fig. 1. The bottom side of the
detector was polished and mirrored 
by an aluminium foil. The other end consists of \linebreak30 cm loose 
fibers. 8 $\times$ 8
fibers of a plane were combined to form a macroscopic pixel. Using a 
plastic connector mask the 16 pixels of a plane were put to a 16      
channel multianode photomultiplier Hamamatsu R5900-M16. The 192 pixels
of the detector are readout in parallel using 12 such devices.

For $\mu FAST$ II both ends of the glued fibers were cut and
polished. This allows to look through the fibers and visually to inspect
their quality. A corresponding photo is shown in fig. 2. The readout
of the detector pixels of the same size as in $\mu FAST$ I is done here   
by clear optical light guides. Using \linebreak500 $\mu$m double clad
KURARAY clear fibers exactly the same planes are produced as for
 scintillating  
fibers with the same winding drum. Polished at one end of the about 5 cm
long glued part, the 30 cm long loose ends are again ordered to
macro-pixels of $8 \times 8$  fibers feed in masks fitting to the same
multianode photomultipliers as mentioned above. Only  
four planes of this type were available  and glued together to a  
compact block. Using pins and holes this block could be connected with
moderate precision to all parts of the $\mu FAST$ II scintillator block.

During the test run the FAST DAQ system has been used to study light   
signals from particles hitting $\mu FAST$ I or II. The signals from the
192 photomultiplier channels were splitted passively. After a beam
trigger the arrival time of all signals appearing in an interval of  
$\pm$ 20 $\mu sec$ were registrated by two VME-TDC's CAEN V767. In addition
42 PM channels were connected to the channels of 
six VME ADC's  LECROY 1182 to measure signal amplitudes within a gate
of 30 nsec after the beam trigger. The detector planes were arranged
perpendicular to the incoming beam defining the z-direction. The
pixels in a plane have increasing row numbers in
y-direction. ADC-channels were connected to all planes of row numbers 8
and 9 and planes 4-12 in row numbers 7 and 10 (see also fig. 2).

At DESY Zeuthen $\mu FAST$ II has been studied during three months after   
the test run using a cosmic ray trigger to activate a VME ADC CAEN V265
for registration of the PM-signal of particular pixels of the detector.

\section{Measurements at PSI}
The basic idea for the irradiation test was to stop an intense pion   
beam in the center of one of the $\mu FAST$ detectors. Due to the strong  
increase of the pion ionization loss near to its stopping point it  
should be possible to irradiate different detector planes with
different but correlated and calculable doses. 

The concept has been tested in a low intensity positive charged
particle beam of the $\pi M1$ area at PSI which is expected to contain mainly
pions. The beam momentum was selected to be 146 MeV/c. The setup is
shown schematically in fig 3a. Beam particles first cross a 8.5 cm   
thick lucite degrader (D) then a beam trigger system of three plastic
scintillators (T1-T3) of 0.5 cm thickness each and hit finally the
planes of the detectors  $\mu FAST$ I or II perpendicularly. Using $\mu
FAST$ I half a million triggers have been recorded for radiation
hardness studies with a data rate of about 100 Hz. 
 
As came out later ~80 $\%$ of all triggers were due to minimum ionizing
particles (positrons) crossing all detector planes. These triggers were used to
calibrate the 12 photomultipliers to give the same signals for any
plane.  The corresponding energy loss spectrum is shown in fig. 4a.

Pions entering the detector were already non-relativistic and gave  
rise to larger energy losses (fig. 4b) increasing to the stopping
point (fig. 4c). Comparing the average values of the three
distributions a ratio of 1.0/2.7/5.6 is found. As required, the pions
stop dominantly in the center of the detector (see fig. 5).

\subsection{Irradiation scheme}
A high intense charged pion beam of the area $\pi E3$
of PSI has been used for a passive irradiation of the $\mu FAST$ II     
scintillator block selecting the same momentum of 146 MeV/c as in $\pi
M1$.  The corresponding arrangement is sketched in 
fig. 3b. The beam crosses a 9.0 cm thick lucite degrader (D). The beam profile
is measured using a transparent wire chamber (C). The beam rate is
monitored by two 0.5 cm thick scintillation counters (T1-T2). Finally
particles enter the detector which was surrounded by a lead shield
otherwise.

Because the scintillation counters could operate only up to moderate
particle rates the PSI proton accelerator intensity was reduced for a
short time by a factor 19.2 to measure the particle flux in our beam
configuration. It was assumed that a linear scaling to the nominal
accelerator intensity is possible also for the $\pi E3$ area. With that   
procedure we measured a particle flux of $0.9 \times 10^8$
particles cm$^{-2} \cdot$ sec$^{-1}$  in a 2 $\times$ 1 cm$^2$ peak region of the beam
profile, illuminating the six central channels (6 - 11) of the detector in
about 5 cm height. Within a total irradiation time of 186000 sec ( 51h
40$^{\prime}$) 1.67 $\times$ 10$^{13}$ particles/cm$^2$ hit that detector region in
total.

To estimate the corresponding radiation dose per detector plane the  
particle content of the beam becomes important. With the available
setup it could not be measured for the conditions in $\pi E3$. In
contrast to the situation in $\pi M1$ where the beam collimators were
nearly closed, they were completely opened here what should decrease positron
appearance.

In fig. 6a the average energy loss in the detector planes as measured
with $\mu FAST I$ in $\pi M1$ is distributed for pions and positrons.
Fig. 6b shows
the part of incoming pions crossing a detector plane or stopping
there. Together with the total particle flux this numbers allow to   
estimate the integral irradiation dose per detector plane. To do that
a GEANT-based Monte Carlo was used to simulate the experimental
conditions and fit the pion stopping distributions in figs. 6b. The
corresponding Monte Carlo result for the dose per plane is displayed in
fig. 6c if the beam consists only of pions or positrons and for the
particle mixture found for the measurement in $\pi$M1. Planes 4 and 5
have been irradiated correspondingly with a dose beetween 1 - 4 Mrad.

\subsection{Irradiation damage}
When the irradiation was finished first a visual inspection of the   
degrader and the $\mu FAST$ II detector was made.

As visible in fig. 7 the degrader showed a clear damage in the region where
the intense beam was crossing. The degrader is made out of two   
blocks pressed strongly together with screws. Both blocks arise from
the same piece of lucite of 5 cm thickness. The second one
was machined down to 4 cm  a few hours before the irradiation started.
During that procedure the block was warmed up considerably. This
circumstance seems to be reflected in an interesting way in the damage
profile. Beam particle hitting the blocks from left produce in the   
first one a brown zone which becomes smaller and weaker near to its
end. At the entrance of the second block however a kind of phase transition
seems to happen. Much stronger browning is observed. The widening of
the profile is due to increasing multiple scattering of pions slowing 
down. The damage seems to be irreversible. After three months no
change of the transparency has been observed.

In contrast to the strong effect for the degrader no visible change of
the transparency of the $\mu FAST$ II detector fibers was seen.

The detector has been studied in addition to $\mu FAST$ I before and after
irradiation in the $\pi M1$ particle beam. However only planes 2-5 were
readout with the optical light guide planes described in section 2. In
table 1 the ratio $R_1$ of the average values of the ADC-spectra for this
planes measured with $\mu FAST$ II and I is given. A light loss of about
20 $\%$ is observed for the clear light guides due to connector coupling
losses because no fiber-by-fiber coupling is possible.

The ratio $R_2$ of the average values of ADC-spectra measured with $\mu FAST$
II after and before the irradiation is also shown in the table. It
comes out to be near to unity, demonstrating that no clear radiation   
damage is appearing for the considered detector region. One would,  
however, like to prove whether the shape of the corresponding spectra
remains unchanged too. That is shown in fig. 8, where the ratio $R_3$ of  
the two sum spectra is distributed for the region of reasonable
statistics of the data. All values are consistent with one within their
errors, giving an average $<R_3> $ which allows a decrease of the observed
light signal of less than 10 $\%$ due to irradiation.

\section{Laboratory measurements with cosmic rays}
To check the long term behavior of the irradiated $\mu FAST$ II detector
for all its twelve planes, the detector was studied using a
cosmic ray particle trigger in the DESY Zeuthen laboratory during
three months. The detector planes
were placed perpendicular to the zenith axis in a black box. On top of the
first plane a scintillation counter allowed to measure at different  
positions along the fibers. The complete trigger was made by a
threefold coincidence of this counter and signals from two planes
directly before and/or after the measured one. For every plane pixel 9
was selected for trigger and measurement. To study any plane of 
$\mu FAST$ II, the light guide block step by step was moved across the
planes. Repeated measurements of the same pixel were done therefore
often with different photomultipliers. For this purpose the 
 PM's were calibrated  using a constant LED light signal.

In fig. 9 the results of all measurements are distributed versus the 
plane number and a rough order in time. In contrast to the impression
from the first measurement no dependence on the
plane number is observed. As can be seen from
fig. 10 all data points follow a gaussian distribution.

Due to lack of time, only a single measurement has been
done before irradiation in plane 11. It fits to the gaussian behavior
of all data.

In fig. 11 the average values of the measurements per plane are shown
together with a total average and its one and three sigma region. The
single data point from before irradiation agrees with that average in
1.4 $\sigma$.

\section{Summary}

A low energy high rate beam of positive pions has been stopped in the
center of a fiber detector made out of 12 planes with 16 pixels each.
The fibers are glued together with white acrylic paint. Every pixel
consists of $8 \times 8$ KURARAY SCSF-78M fibers of 500 $\mu$m diameter.

Within about 52 hours a total dose of 1 - 4 Mrad has been placed in
the detector center with a rate of 20 - 80 krad/h. The large error of
this number is due to a unknown positron contribution to the
beam.

Unexpectedly, all  measurements are consistent with no radiation   
damage of any fiber detector plane. However, a large and
irreversible damage (browning) reflecting the profile of the crossing
beam is observed for the lucite degrader placed in front of the detector.
Within the measurement errors we can not exclude a maximum
decrease of 10 $\%$ of the light signals of detector fibers after
irradiation. If present, no recovery has been observed during three
months.

\section*{Acknowledgement}

First of all we would like to thank the FAST collaboration members
from BNL, Bologna, CERN, PSI and ETH Zurich who heavily supported the
radiation hardness measurements. Secondly the whole program would not
have been possible without tremendous help provided by the Paul
Scherrer Institute. In particular we thank C. Petitjean, K. Deiters,
and J. Egger for support in the installation of the necessary hardware
and the accelerator crew for providing the required intensities for
monitoring and exposure.

\newpage

\begin{center}
   Table 1 \\

\vspace*{1cm}
\begin{tabular}{|c|c|c|c|} \hline \hline
 Plane & $R_1$ &  $R_2$ &  $ < R_3 > $     \\ \hline \hline
  2    & 0.81  &  1.04  &  0.98 $\pm$ 0.05 \\ \hline          
  3    & 0.75  &  0.94  &  0.93 $\pm$ 0.05 \\ \hline
  4    & 0.78  &  1.07  &  0.96 $\pm$ 0.03 \\ \hline
  5    & 0.85  &  1.07  &  0.94 $\pm$ 0.03 \\ \hline
 sum   & 0.81  &  1.04  &  0.94 $\pm$ 0.02 \\ \hline \hline
\end{tabular} \\

\end{center}

For the detector planes 2-5 and their sum three ratios of ADC spectra
are given:  $R_1$ -the ratio of average values of spectra measured for
$\mu FAST$ II and $\mu FAST$ I, - $R_2$  - the ratio of average values of
the spectra measured for $\mu FAST$ II after and before irradiation - and
$<R_3>$ - the average value of the ratio of ADC-channel contents measured
after and before irradiation.

\newpage

{\large \bf Figure captions}\\ 

\begin{tabular}{lp{12cm}}
Fig. 1 : & Photo of the installation of the $\mu FAST$ I detector within a
     black box. The loose ends of the scintillating fibers are coupled
     via plastic masks to photomultipliers which could be installed on
     top of the box.\\  
Fig. 2 : & Photo of a DESY-logo through the 10 cm long fibers of the $\mu
     FAST$ II detector. The fiber planes and pixels in z and y-direction are indicated as  well as the region in the center where pulse height measurements were  possible \\
Fig. 3 : & Sketch of the beam setups for a.) quantitative studies of
the performance of $\mu FAST$ I and II,  b.) irradiation,  D :
degrader, T1-T3: trigger counter, C: wire chamber. \\
Fig. 4 : & Energy loss distributions for a.) positrons crossing $\mu FAST$ I,
         b.) pions in the first plane, c.) pions in the stopping point. \\
Fig. 5 : & Distribution of pion stoppings per detector plane of $\mu FAST$  I. \\
Fig. 6 : & a.) average energy loss as measured for pions and positrons
per plane of $\mu FAST$ I, b.) corresponding fraction of crossing and
stopping pions, c.) results of a GEANT based Monte Carlo calculation of the radiation dose placed per  plane for different assumptions about the particle content  of the incoming beam. \\
Fig. 7 : & Photo of the side view of the degrader after irradiation  with clearly visible damage in the irradiation region. The beam particles are entering from the left.                  \\
Fig. 8 : & Ratio of the sum of ADC-spectra for planes 2-5 of $\mu FAST$ II
 after and before irradiation. \\
Fig. 9 : & Mean ADC-values as measured for cosmic ray particle crossing 
 a pixel of a certain plane of $\mu FAST$ II   \\
Fig. 10 : & Number distribution of measurements displayed in fig. 9 
versus the measured variable. \\
Fig. 11 :&  Average of mean ADC-values for cosmic ray particle
crossings  per plane of $\mu FAST$ II. \\
\end{tabular}

\newpage
\begin{figure*}[b]
\vspace*{-0.5cm}
\begin{center}
\epsfig{file=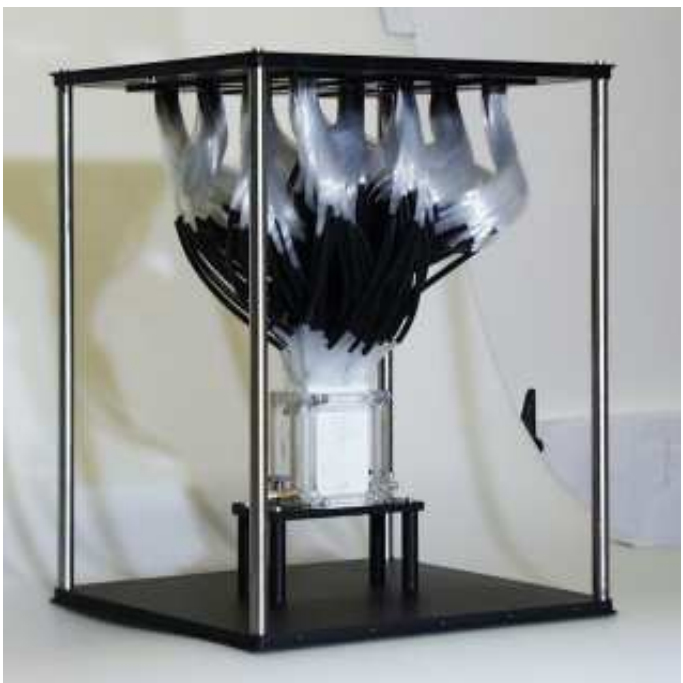,width=9cm,height=9cm
}
\caption{\label{fig1} Photo of the installation of the $\mu FAST$ I
 detector within a black box. The loose ends of the scintillating fibers
 are coupled via plastic masks to photomultipliers which could
 be installed on top of the box.}
\end{center}
\end{figure*}
\begin{figure*}[b]
\vspace*{-0.5cm}
\begin{center}
\epsfig{file=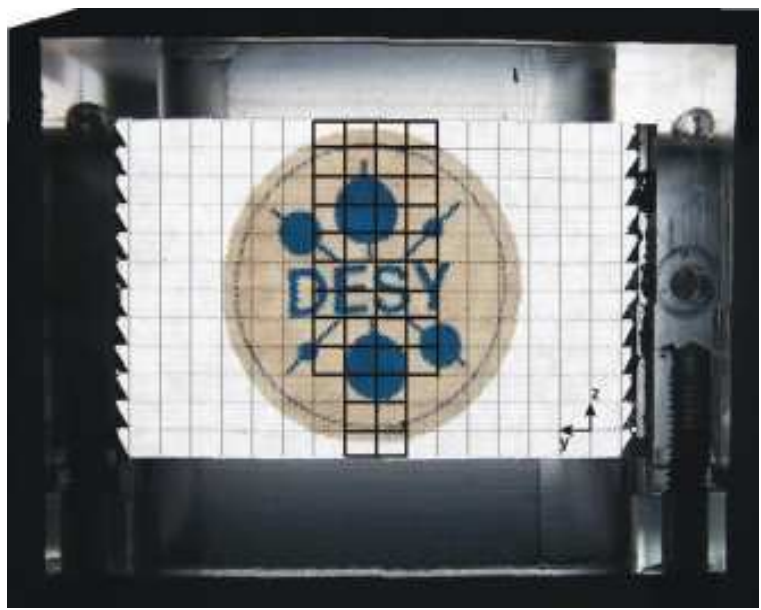,width=10cm,height=8cm}
\caption{\label{fig2} Photo of a DESY-logo through the 10 cm long fibers
 of the $\mu FAST$ II detector.The fiber planes and pixels in z and
 y-direction are indicated as well as the region in the center where
 pulse height measurements were possible }
\end{center}
\end{figure*}
\begin{figure*}[b]
\vspace*{-0.5cm}
\begin{center}
\epsfig{file=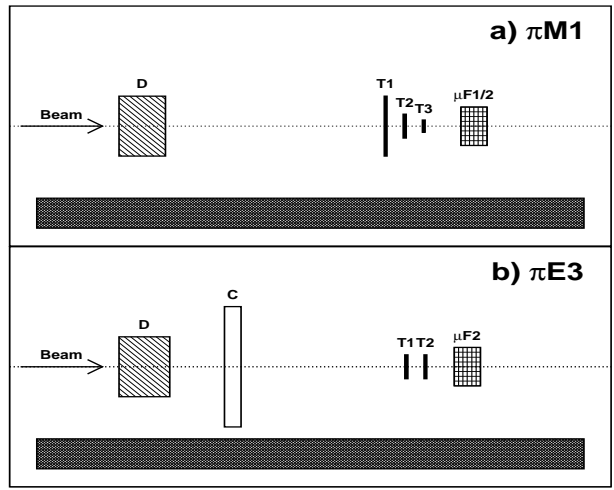,width=10cm,height=8cm}
\caption {\label{Fig.3} : Sketch of the beam setups for a.) quantitative
studies of the performance of $\mu FAST$ I and II, b.) irradiation
D : degrader, T1-T3: trigger counter, C: wire chamber.}  
\end{center}
\end{figure*}
\begin{figure*}[b]
\vspace*{-0.5cm}
\begin{center}
\epsfig{file=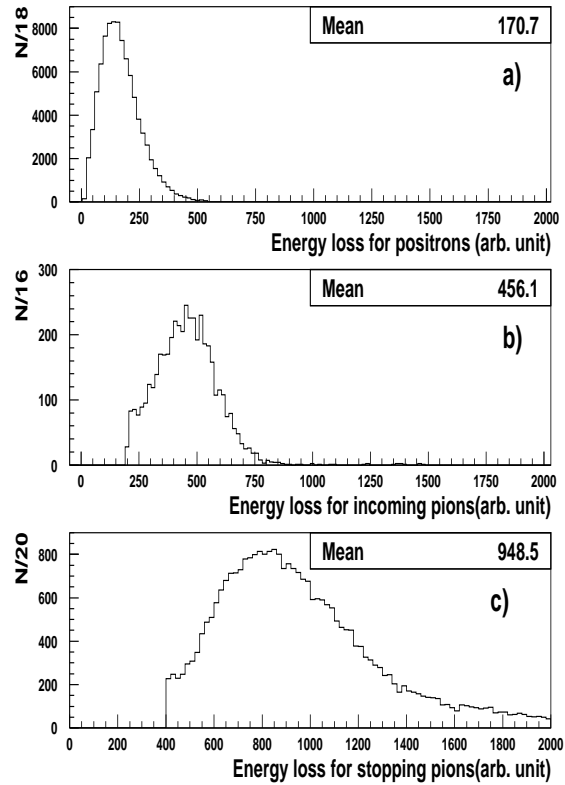,width=8cm,height=12cm
}
\caption {\label{Fig.4}  Energy loss distributions for a.) positrons
 crossing $\mu FAST$ I, b.) pions in the first plane, c.) pions in the
 stopping point }
\end{center}
\end{figure*}   
\begin{figure*}[b]
\vspace*{-0.5cm}
\begin{center}
\epsfig{file=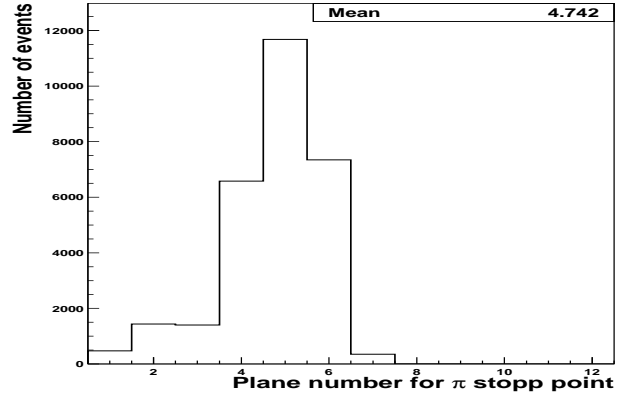,width=10cm,height=6cm
}
\caption {\label{Fig.5} Distribution of pion stoppings per detector plane
of $\mu FAST$  I. }
\end{center}
\end{figure*}
\begin{figure*}[b]
\vspace*{-0.5cm}
\begin{center}
\epsfig{file=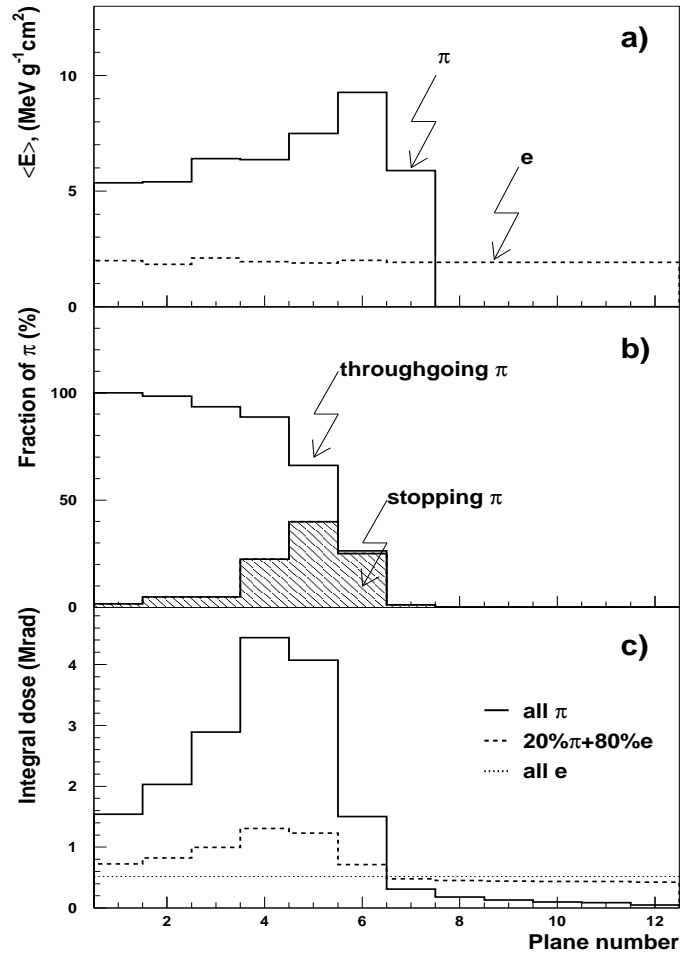,width=10cm,height=14cm}
\vspace*{-0.5cm}
\caption {\label{Fig.6 }  a.) average energy loss as measured for pions
 and positrons per plane of $\mu FAST$ I, b.) corresponding fraction of   
 crossing and stopping pions, c.) radiation dose placed per     
 plane for different assumptions about the particle content  
 of the incoming beam. }                                       
\end{center}
\end{figure*}
\begin{figure*}[b]
\vspace*{-0.5cm}
\begin{center}
\epsfig{file=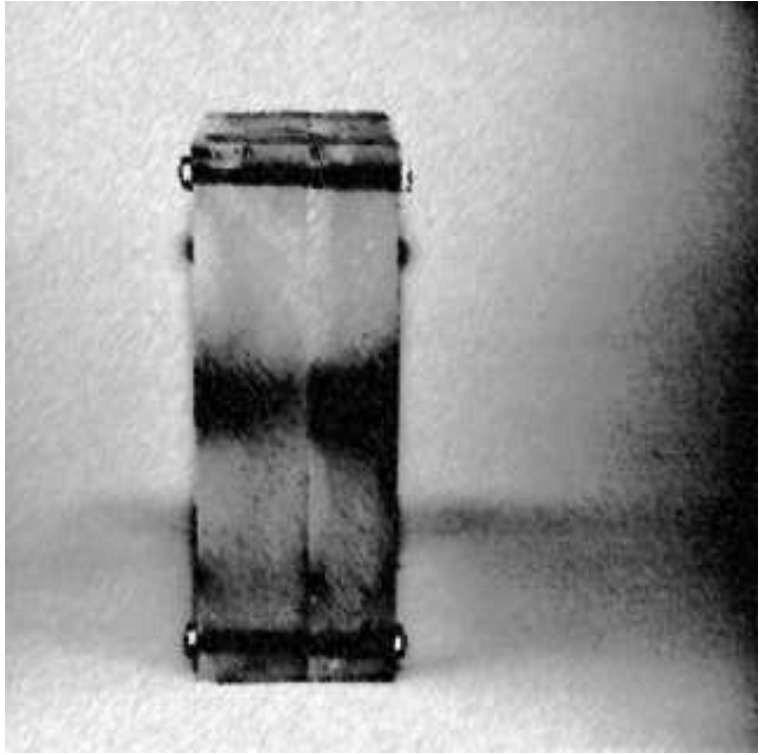,width=10cm,height=10cm}
\caption{\label{fig7} Photo of the side view of the degrader
 after irradiation with clearly visible damage in the irradiation
 region. The beam particles are entering from the left. }
\end{center}
\end{figure*}
\begin{figure*}[b]
\vspace*{-1.5cm}   
\begin{center}
\epsfig{file=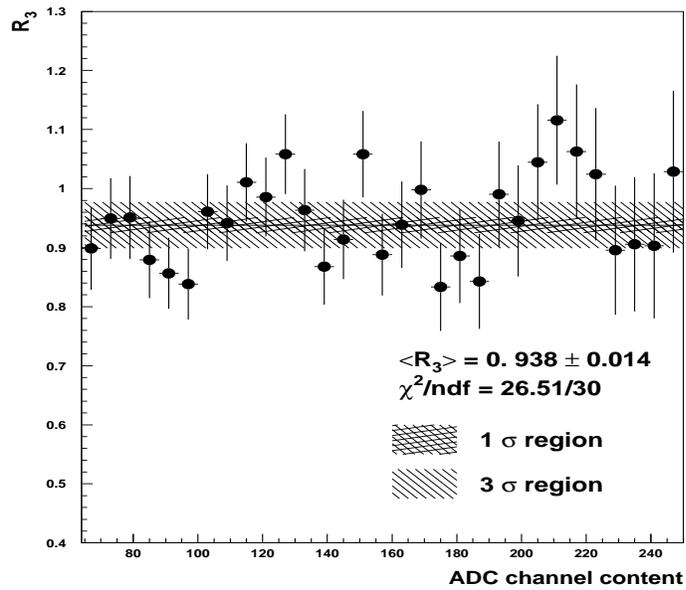,width=10cm,height=11cm 
}
\vspace*{-0.5cm}
\caption {\label{Fig.8 } Ratio of the sum of ADC-spectra for planes 2-5 of
$\mu FAST$ II after and before irradiation. }
\end{center}
\end{figure*}
\begin{figure*}[b]
\vspace*{-0.5cm}
\begin{center}
\epsfig{file=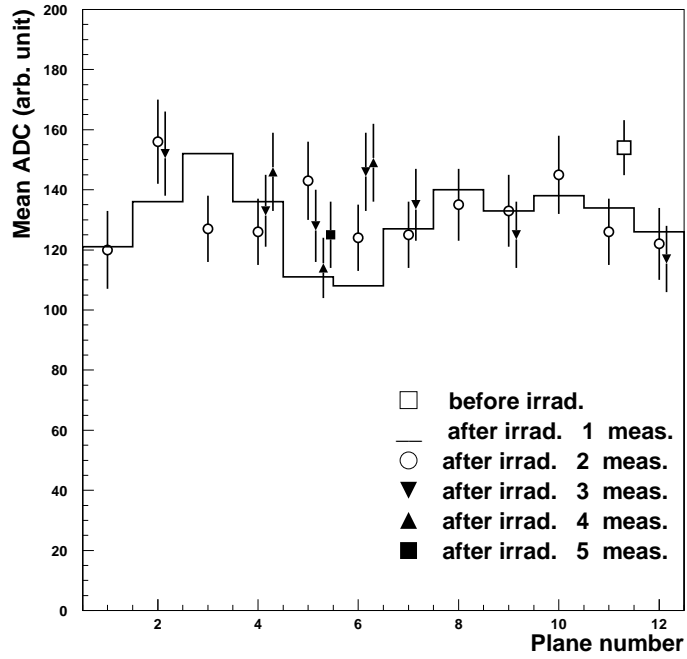,width=10 cm }
\caption {\label{Fig.9 } Mean ADC-values as measured for cosmic ray
 particle crossing a pixel of a certain plane of $\mu FAST$ II }  
\end{center}
\end{figure*}   
\begin{figure*}[b]
\vspace*{-0.5cm} 
\begin{center}
\epsfig{file=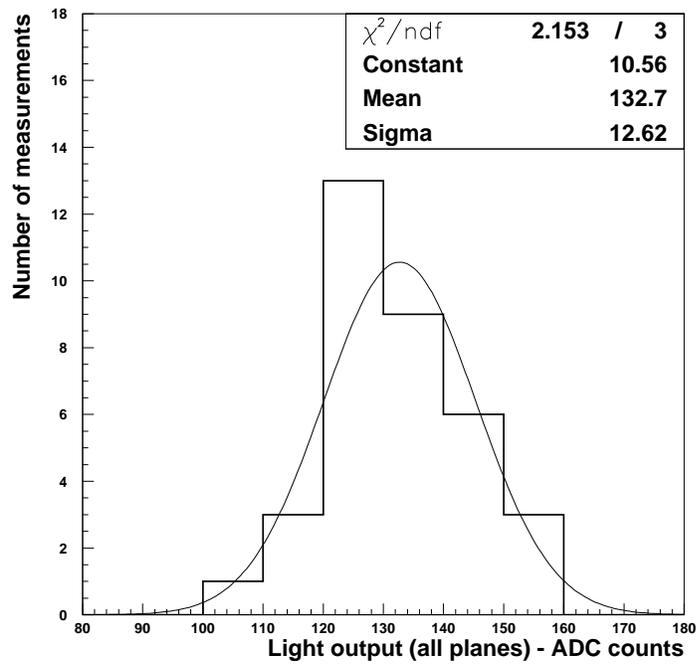,width=10 cm }
\caption {\label{Fig.10} Number distribution of measurements displayed in
fig. 9 versus the measured variable. }
\end{center}
\end{figure*}   
\begin{figure*}[b]
\vspace*{-0.5cm}
\begin{center}
\epsfig{file=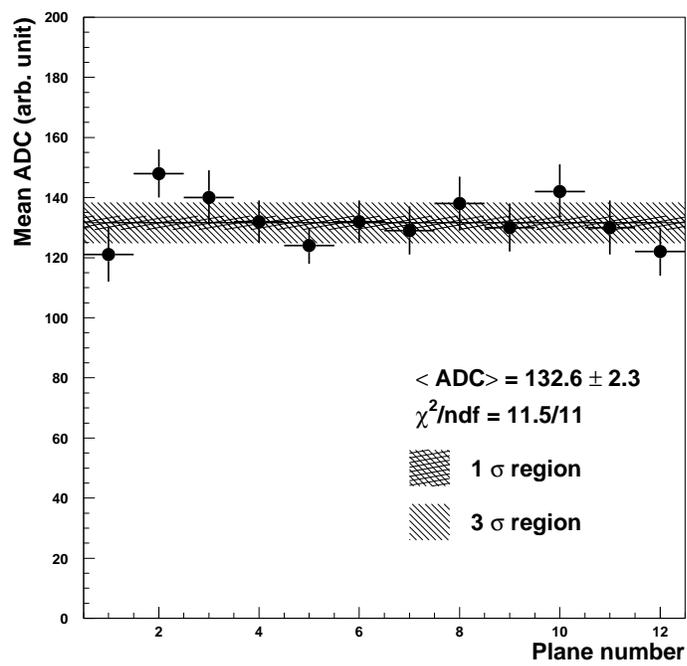,width=10 cm }
\caption {\label{Fig.11}  Average of mean ADC-values for cosmic ray
particle crossings per plane of $\mu FAST$ II.}

\end{center}
\end{figure*}

\end{document}